\documentclass[prc,showpacs,twocolumn,floatfix,superscriptaddress]{revtex4}
\usepackage{graphicx}
\usepackage{amssymb}
\usepackage{amsmath}
\usepackage{dcolumn}
\usepackage{bm}
\renewcommand{\vec}[1]{\bm{\mathrm{#1}}}
\newcommand{\HF}{\textit{HF}}
\newcommand{\scatt}{\textit{scatt}}
\newcommand{\bound}{\textit{bound}}
\newcommand{\corr}{\textit{corr}}
\newcommand{\Mott}{\textit{Mott}}
\newcommand{\tot}{\textit{tot}}
\newcommand{\liqgas}{\textit{liq-gas}}
\DeclareMathOperator{\Real}{Re}
\DeclareMathOperator{\Imag}{Im}
\renewcommand{\Re}{\Real}
\renewcommand{\Im}{\Imag}
\newcommand{\Eq}[1]{Eq. (\ref{#1})}
\newcommand{\Eqs}[1]{Eqs. (\ref{#1})}
\newcommand{\Fig}[1]{Fig. \ref{#1}}
\newcommand{\Figs}[1]{Figs. \ref{#1}}
\newcommand{\Ref}[1]{Ref. \cite{#1}}
\newcommand{\Refs}[1]{Refs. \cite{#1}}
\newcommand{\Sec}[1]{Sec. \ref{#1}}

\begin{document}

\title{BEC-BCS Crossover and the Liquid-Gas Phase Transition
in Hot and Dense Nuclear Matter}
\author{Meng Jin}
\affiliation{Institute of Particle Physics and Physics Department,
Hua-Zhong Normal University, Wuhan 4300079, People's Republic of
China}
\author{Michael Urban}
\affiliation{Institut de Physique Nucl\'eaire, CNRS/IN2P3 and
Universit\'e Paris-Sud 11, 91406 Orsay Cedex, France}
\author{Peter Schuck}
\affiliation{Institut de Physique Nucl\'eaire, CNRS/IN2P3 and
Universit\'e Paris-Sud 11, 91406 Orsay Cedex, France}
\affiliation{Laboratoire de Physique et Mod\'elisation des Milieux
Condens\'es, CNRS and Universit\'e Joseph Fourier, BP 166, 38042
Grenoble Cedex, France}
\begin{abstract}
The effect of nucleon-nucleon correlations in symmetric nuclear matter
at finite temperature is studied beyond BCS theory. Starting from a
Hartree-Fock description of nuclear matter with the Gogny effective
interaction, we add correlations corresponding to the formation of
preformed pairs and scattering states above the superfluid critical
temperature within the in-medium $T$-matrix approach, which is
analogous to the Nozi\`eres-Schmitt-Rink theory. We calculate the
critical temperature for a BEC superfluid of deuterons, of a BCS
superfluid of nucleons, and in the crossover between these limits. The
effect of the correlations on thermodynamic properties (equation of
state, energy, entropy) and the liquid-gas phase transition is
discussed. Our results show that nucleon-nucleon correlations beyond
BCS play an important role for the properties of nuclear matter,
especially in the low-density region.
\end{abstract}
\date{May 11, 2010}
 \maketitle

\section {Introduction}
Pairing and nucleon-nucleon correlations are important properties of
interacting nuclear systems. For example, in the weak-coupling limit,
i.e., at high density, the nucleons form Cooper pairs, and below a
certain critical temperature $T_c$ the system is in a superfluid
phase as described by the Bardeen-Cooper-Schrieffer (BCS) theory. In
the strong-coupling limit, i.e., at low density, neutrons and protons
form deuteron bound states which will condense if the temperature is
below the critical temperature for the corresponding Bose-Einstein
condensation (BEC). It was theoretically predicted \cite{nsr} and
recently confirmed by experiments with ultracold atomic Fermi gases
\cite{regal,zwierlein} that there is a smooth crossover between the
BCS and BEC limits. Qualitatively, especially at zero temperature,
these features can be studied within the BCS (mean field)
approximation \cite{baldo}. Quantitatively, however, the critical
temperature obtained in this way is too high because the BCS theory
does not include the existence of non-condensed pairs at finite
temperature. In order to go beyond mean field, one has to consider
pair correlations already above the critical temperature, as in the
Nozi\`eres-Schmitt-Rink (NSR) theory \cite{nsr}. Especially in the low
density region, where the coupling between nucleons is strong, such
correlations modify the mean-field results to a large extent.

At present, there are several groups who have studied nuclear matter
within the NSR approach. Pioneering work has been done by the
Rostock group \cite{schmidt,stein_nsr}. There are also extensions
where the correlations are considered in a more self-consistent way,
like in the self-consistent Green's function method
\cite{rios,soma}. A generalization to temperatures below the
superfluid transition temperature was discussed by Bo\.zek
\cite{bozek}. In the case of ultracold Fermi gases, where the results
can be compared with very precise measurements, theories for the
BEC-BCS crossover based on the NSR approach \cite{perali} have been
very successful \cite{pieri}.

It is well known that there exists a liquid-gas phase transition in
nuclear matter. Experimental information can be obtained from
multifragmentation (see, e.g., \cite{chomaz,borderie,ma,moretto}). The
critical temperature deduced from these experiments depends on the
mass of the nuclei and can be as low as 6.7 MeV \cite{moretto} in the
case of small systems. For infinite nuclear matter, theoretical
predictions give much higher values for the critical temperature
between 14 and 18 MeV \cite{chomaz,borderie} (see \Ref{typel} for a
recent theoretical study). Below that temperature, nuclear matter is
unstable in a certain range of low densities. Within mean-field
theory, we know that the BCS-BEC crossover is completely covered by
the instability region of the liquid-gas phase
transition. Nevertheless, the investigation of low-density nuclear
matter is of interest for applications where regions of low density
appear in the framework of the local-density approximation. Contrary
to the nuclear matter case, the whole crossover can be studied in the
case of ultracold atomic Fermi gases \cite{regal,zwierlein}, because
the pair correlations stabilize the gas \cite{nsr} such that the
system does not collapse into its solid ground state but it remains in
its metastable gas state. By analogy, one expects that pair
correlations will stabilize low-density nuclear matter and thus reduce
the liquid-gas coexistence region. One of our subjects of
investigation will be how strong this effect of nucleon-nucleon
correlations on the liquid gas phase transition is quantitatively.

Furthermore, in this paper we will calculate the equation of state of
hot and dense symmetric nuclear matter, taking into account the
contribution of the mean field together with the nucleon-nucleon
correlations. For the mean field we will use the Gogny interaction
because it is known to give a good description of the single-particle
and thermodynamic properties of nuclear matter, including saturation
at the right density, the liquid-gas phase transition, etc. For the
part beyond the mean field, we use the $T$-matrix (or ladder
approximation) which contains the information on two-particle
correlations. This also allows us to extract the critical temperature
for pair condensation smoothly interpolating between the BEC and BCS
regimes.

The paper is organized as follows. In \Sec{sec:formalism}, we will give a
summary of the formalism. The numerical results are provided in
\Sec{sec:numerics}. The last section is devoted to the summary and
discussions.
\section {Formalism}
\label{sec:formalism}
Before explicitly including two-particle correlations, we calculate
the single-particle Green's function within the Hartree-Fock (HF)
approximation. In order to get a reasonable description of the
single-particle energies, we use the density-dependent D1 Gogny
effective interaction to describe the mean field. This force gives
nuclear binding at the right saturation point and many other
properties of nuclear matter and of finite nuclei \cite{gogny}. It has
the form
\begin{multline}
V(r) = \sum_{m=1}^2(W_m+B_m P_\sigma-H_m P_\tau-M_mP_\sigma P_\tau)
  e^{-r^2/\mu_m^2}\\ +t_0(1+x_0P_\sigma)\rho^\alpha\delta(r),
\end{multline}
where the $P_\sigma$ and $P_\tau$ are, respectively, the spin and
isospin exchange operators. The spin-orbit coupling term is neglected
here, since we consider only the properties of infinite nuclear
matter. For the parameters we use the values given in \Ref{gogny}
\footnote{We prefer the D1 parametrization to the D1S one
\cite{berger} because it allows us to compare our HF results with
those of \Ref{ventura} and it gives a better compressibility of
symmetric nuclear matter \cite{margueron}. Anyway, since the effective
mass $m^*$ in D1 and D1S is almost the same, the results do not change
qualitatively if we use D1S instead of D1.}.
For details of the HF description of nuclear matter at finite
temperature with the Gogny force, see \Refs{heyer,song,ventura}. The
HF mean field $\Sigma_\HF$ contains the direct, the exchange, and the
rearrangement contributions. Because of the finite range of the Gogny
force, the exchange contribution is momentum dependent, and the
single-particle Green's function takes the form
\begin{equation}
G_\HF(p,\omega) = \frac{1}{\omega-\xi_p+i0}\,,
\end{equation}
where $\xi_p$ is the quasiparticle energy defined by
\begin{equation}
\xi_p = \frac{p^2}{2m}-\Sigma_\HF(p)-\mu\,,
\label{xiHF}
\end{equation}
where $\mu$ denotes the chemical potential. In order to facilitate the
numerical calculation of the correlation effects, we use the
effective-mass approximation for the Gogny mean field, i.e., we write
\cite{heyer}
\begin{equation}
\xi_p = \frac{p^2}{2m^*}-\mu^*\,.
\label{ximstar}
\end{equation}
There are different ways to define the effective nucleon mass
$m^*$. In principle, $m^*$ is momentum dependent \cite{ventura}. Here
we use the effective mass defined by expanding \Eq{xiHF} around
$p=0$ (we checked that for the final results it makes almost no
difference if we expand around zero or around the Fermi momentum),
i.e.,
\begin{gather}
\frac{1}{m^*}=\frac{1}{m}+2\left.\frac{d\Sigma_\HF(p)}{dp^2}
  \right|_{p=0},\\
\mu^*=\mu-\Sigma_\HF(0)\,.
\end{gather}
However, the effective-mass approximation will only be used for the
calculation of the correlation effects, while the mean-field
contributions will be computed with the full momentum dependence of
$\Sigma_\HF(p)$.

\begin{figure}
\includegraphics[width=8.5cm,clip=true]{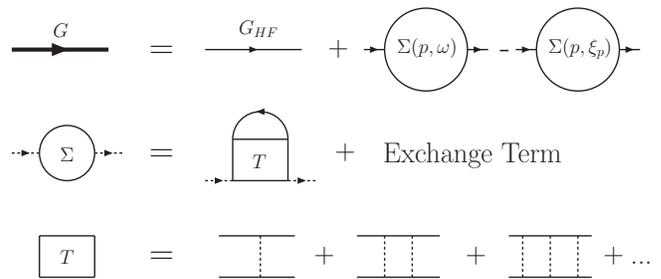}
\caption{The Feynman diagrams for the Green's function (top), for the
self-energy (middle), and for the $T$-matrix in ladder approximation
(bottom).
\label{fig:feynman}}
\end{figure}

In principle, we are looking for the full single-particle Green's
function $G$ including correlations. The Dyson equation can be written
as
\begin{equation}
G^{-1}(p,\omega) = G^{-1}_\HF(p,\omega)-\tilde{\Sigma}(p,\omega)\,,
\label{Dysoneq}
\end{equation}
where $\tilde{\Sigma}$ is the correlation contribution to the
single-particle self-energy. Since the Gogny force is a
density-dependent effective interaction, which is designed to give
good results already at the HF level, we suppose that the Gogny mean
field accounts already for most of the correlation effects. We
therefore demand that the correlations do not shift the quasiparticle
energies $\xi_p$, i.e., $\tilde{\Sigma}(p,\xi_p) = 0$, and that the
role of the correlations is just to reduce the strength of the
quasiparticle pole and to distribute the remaining strength in the
continuum of the spectral function. Hence, we define $\tilde{\Sigma}$
to be the self-energy subtracted at $\xi_p$:
\begin{equation}
\tilde{\Sigma}(p,\omega) = \Sigma(p,\omega) - \Re \Sigma(p,\xi_p)\,.
\label{Sigmasub}
\end{equation}
In order to describe pair correlations, we calculate the self-energy
$\Sigma$ within the $T$-matrix or ladder approximation, as shown in
the lower part of \Fig{fig:feynman}.
This is a frequently used lowest-order correction
\cite{nsr,schmidt,stein_nsr,perali}, implying, however, that vertex
corrections as well as screening of the interaction due to the medium
effects are neglected.

Since our aim is not a completely self-consistent description of the
spectral function as in the self-consistent Green's function method
\cite{rios,soma}, we make the assumption that the correlations can be
treated as a small correction to the Gogny HF self-energy. This allows
us to use the HF Green's function $G_\HF$ in the calculation of the
$T$ matrix and of the self-energy $\Sigma$. Then, for consistency,
one should also keep only the first-order term of \Eq{Dysoneq}, i.e.,
\begin{equation}
G(p,\omega)= G_\HF(p,\omega)+G_\HF^2(p,\omega)\tilde{\Sigma}(p,\omega)\,.
\label{Gcorr}
\end{equation}
A diagrammatic representation is given in the upper part of
\Fig{fig:feynman}

That the self-energy in T-matrix approximation should only be treated
in first-order perturbation theory may also have a more formal
reason. The T-matrix approximation corresponds to particle-particle
random-phase approximation (pp-RPA) \cite{rs}.  It can be shown that
the ground-state energy calculated from the single-particle Green's
function with self-energy in first order and in T-matrix approximation
yields exactly the pp-RPA ground-state energy \cite{bouyssy}. At least
this holds true for the self-energy without subtraction
procedure. Therefore our formalism is closely related to that of
Ref. \cite{jiang}, where the pp-RPA formalism is used, except that we
apply the subtraction prescription while the authors of
Ref. \cite{jiang} are obliged to reduce the correlation contribution
by introducing a cutoff and to change the parameters of the Gogny
force in order maintain the right saturation point of nuclear matter.

Note that our approximations are analogous to NSR theory \cite{nsr},
except that in NSR theory free Green's functions instead of HF ones
are used and consequently no subtraction is made in the
self-energy. In the case of nuclear matter, however, we cannot expect
to obtain a good description of the full self-energy from such a
simple model for the $T$ matrix. This is why we use the Gogny mean
field and the subtraction method described above, while the subtracted
self-energy serves only to provide the energy dependence corresponding
to the pair correlations in the channels we want to study.

\begin{figure}
\includegraphics[width=7.5cm]{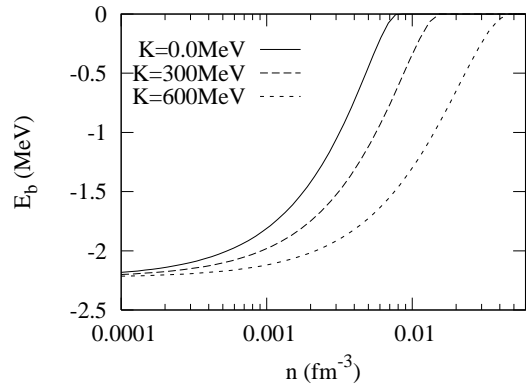}
\caption{The deuteron binding energy in nuclear matter from the
  Yamaguchi potential and including the effect of the Gogny mean
  field, as a function of the density for different values of the
  deuteron momentum $K$. The temperature is $T = 10$
  MeV.\label{fig:binding}}
\end{figure}

\begin{figure}
\includegraphics[width=7.5cm]{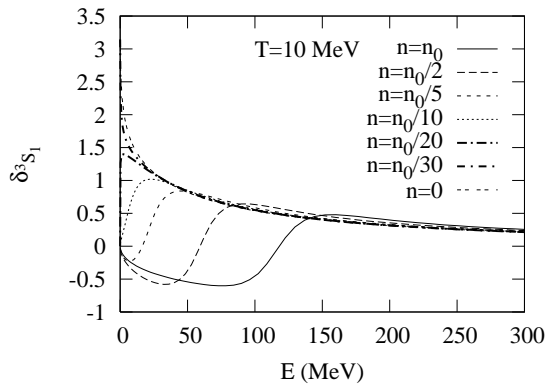}
\caption{In-medium scattering phase shift in the $^3S_1$ channel for
$K=0$ as a function of $E = k^2/m^*$ for different densities and $T =
10$ MeV.\label{fig:phaseshift}}
\end{figure}

In order to get a simple expression for the $T$ matrix, we use the
separable Yamaguchi potential \cite{yamaguchi},
\begin{equation}
V_\alpha(k,k') = -\lambda_\alpha v(k)v(k')
\end{equation}
where $k$ and $k'$ are the incoming and outgoing relative momenta in
the center-of-mass frame, and the form factor is given by
\begin{equation}
v(k)= \frac{1}{k^2+\beta^2}.
\end{equation}
As in \Ref{stein_nsr}, we consider only S-wave scattering ($\alpha=$
$^1S_0$, $^3S_1$) and neglect the coupling between the $^3S_1$ and
$^3D_1$ channels (which comes from the tensor force).  With the
parameters $\beta=1.4488$ fm$^{-1}$, $\lambda_{^1S_0}=2994$ MeV
fm$^{-1}$ and $\lambda_{^3S_1}=4264$ MeV fm$^{-1}$ \cite{stein_nsr},
the low-energy nucleon-nucleon phase shifts and the vacuum binding
energy of the deuteron ($E_b^0=-2.225$ MeV) are very well reproduced,
see results for $n=0$ in \Figs{fig:binding} and \ref{fig:phaseshift},
so that it is unlikely that the coupling between the $^3D_1$ and
$^3S_1$ channels would strongly modify our results.
With the separable interaction, the resummation of the ladder diagrams
shown in the lower part of \Fig{fig:feynman} reduces to a simple
geometrical series, and the $T$ matrix can be written as
\begin{equation}
T_\alpha(k,k',K,\omega)=\frac{V_\alpha(k,k')}{1-J_\alpha(K,\omega)}\,,
\label{Tmatrix}
\end{equation}
where $\vec{k}$ and $\vec{k}'$ are the incoming and outgoing momenta
in the center of mass frame, $\vec{K}$ is the total momentum, and
\begin{multline}
J_\alpha(K,\omega)=\int\frac{d^3k}{(2\pi)^3}V_\alpha(k,k)\\
  \times \frac{1-f(\xi_{\vec{K}/2+\vec{k}})-f(\xi_{\vec{K}/2-\vec{k}})}
    {\omega-\xi_{\vec{K}/2+\vec{k}}-\xi_{\vec{K}/2-\vec{k}}+i0}\,.
\label{Jalpha}
\end{multline}
The function $f(\xi) = 1/(e^{\xi/T}+1)$ is the Fermi function, $T$
being the temperature. Within the effective mass approximation,
\Eq{ximstar}, the denominator of \Eq{Jalpha} does not depend on the
angle between $\vec{k}$ and $\vec{K}$, and the angular integral can be
done analytically. The main contribution to the integral over the
relative momentum comes from low momenta due to the form factor of the
Yamaguchi interaction ($k\lesssim\beta$).

In the $^3S_1$ channel, it can happen that $J_{^3S_1}(K,\omega_b) = 1$
at some energy $\omega_b$ below the threshold energy
\begin{equation}
\omega_0(K) = \frac{K^2}{4m^*}-2\mu^*\,.
\end{equation}
This means that there is a bound state (the deuteron) with binding
energy $E_b(K) = \omega_b(K)-\omega_0(K)$. As an example, the deuteron
binding energies for different values of the deuteron momentum $K$ are
displayed in \Fig{fig:binding}. As one can see, the binding gets
weaker with increasing density, and eventually the deuteron gets
unbound at the so-called Mott density. Since the Pauli blocking effect
gets weaker with higher deuteron momentum $K$, there exists for any
density a Mott momentum $K_\Mott$ above which the deuteron stays
bound.

The in-medium nucleon-nucleon phase shifts $\delta_\alpha$ can easily
be obtained from $1/(1-J_\alpha) = e^{i\delta_\alpha} /
|1-J_\alpha|$. As an example, we show in \Fig{fig:phaseshift} the
phase shift in the $^3S_1$ channel for $K=0$ at different densities,
as function of the energy $E = \omega+2\mu^* = k^2/m^*$. We see that
at higher densities, e.g., at $n \geq n_0/5$ ($n_0 = 0.17$ fm$^{-3}$
being the saturation density of nuclear matter), the phase shift is
negative in the low-energy region and then becomes positive as the
energy increases. The energy where the phase shift crosses zero is
$\omega = 0$, i.e., $E = 2\mu^*$. At lower densities, when $\mu^*$ is
negative, the phase shift is positive at low energy. At some very low
density, the value of the phase shift at $E = 0$ changes from 0 to
$\pi$. This happens precisely at the density below which the deuteron
is bound.

In terms of the $T$ matrix, we can write the self-energy $\Sigma$
depicted in the middle of \Fig{fig:feynman} within the Matsubara
formalism as
\begin{multline}
\Sigma(p,i\omega_n) = \tfrac{3}{2} \sum_{\alpha=\,^3S_1,\,^1S_0}
  T \sum_{n'} \int \frac{d^3p'}{(2\pi)^3} G_\HF(p',i\omega_{n'})\\
  \times T_\alpha(k,k,K,i\omega_n+i\omega_{n'})\,,
\label{SigmaMatsubara}
\end{multline}
where $\omega_n$ and $\omega_{n'}$ are Fermionic Matsubara frequencies
[$\omega_n = (2n+1)\pi T$], $\vec{k} = (\vec{p}-\vec{p}')/2$, and
$\vec{K} = \vec{p}+\vec{p}'$. The factor $3/2$ is the product of a
factor $1/4$ from the averaging over spin and isospin in symmetric
nuclear matter, of a factor $2$ from the sum of direct and exchange
contributions, and of a factor $(2S+1)(2T+1) = 3$ for $\alpha = ^3S_1$
and $^1S_0$ from the sum over spin and isospin in the loop. Using
standard techniques \cite{fetter}, the self-energy can be analytically
continued to real energies, which is necessary for the calculation of
the subtraction term $\Sigma(p,\xi_p)$ in \Eq{Sigmasub}.

Inserting the self-energy into \Eq{Gcorr}, we calculate the
density from
\begin{equation}
n(T,\mu) = -4 T\sum_{n} \int \frac{d^3k}{(2\pi)^3} G(k,i\omega_n) \,.
\end{equation}
The factor $4$ comes from the sum over spin and isospin. It is clear
that the first term of \Eq{Gcorr} just gives the Hartree-Fock
density, and the second term gives the correction beyond the mean
field approximation. After a lengthy derivation (see Appendix), one
finds the following formulas initially given in \Refs{schmidt,stein_nsr}:
\begin{equation}
n = n_\HF + n_\corr = n_\HF+n_\bound+n_\scatt\,.
\label{ncontrib}
\end{equation}
The bound-state contribution reads
\begin{equation}
n_\bound = 6 \int_{K > K_\Mott} \frac{d^3K}{(2\pi)^3} g(\omega_b(K))\,,
\label{nbound}
\end{equation}
where $g(\omega) = 1/(e^{\omega/T}-1)$ is the Bose function. This term
gives the nucleon density corresponding to a Bose gas of
deuterons. The scattering-state contribution reads
\begin{multline}
n_\scatt =
-6 \int_{K > K_\Mott}
  \frac{d^3K}{(2\pi)^3} g(\omega_0(K))\\
-6 \sum_{\alpha=\,^3S_1,\,^1S_0}
  \int\frac{d^3K}{(2\pi)^3} \int_{\omega_0(K)}^\infty \frac{d\omega}{\pi}
  \left(\frac{d}{d\omega}g(\omega)\right)\\ \times
  \left(\delta_\alpha-\tfrac{1}{2}\sin 2\delta_\alpha\right).
\label{nscatt}
\end{multline}
In \Ref{schmidt}, these equations were derived in a different way using
the optical theorem, analogously to the derivation of a similar
formula for the electron-hole system in \Ref{zimmermann}.

Note that in spite of the double pole of the derivative of the Bose
function at $\omega = 0$, the integrand in \Eq{nscatt} has no
pole. This is because $\delta_\alpha$ crosses zero at $\omega =
0$. This simple zero is raised to a double one due to the difference
of the two terms in the second line of \Eq{nscatt} \footnote{Note also
that the statement in \Ref{stein_nsr}, saying that \Eq{nscatt} reduces
to the NSR formula for the density after integration by parts if the
term $\propto \sin 2\delta_\alpha$ is omitted, is incorrect. In fact,
the NSR formula involves a derivative $d\delta/d\mu$ instead of
$d\delta/d\omega$ and therefore does not have a pole in the integrand
even if that term is omitted. The term $\propto \sin 2\delta_\alpha$
cannot be identified with the contribution of the subtraction of
$\Sigma(p,\xi_p)$ in \Eq{Sigmasub}.}.

Once we have calculated the density, we can calculate the pressure. To
that end, we integrate the thermodynamic relation $n = (dP/d\mu)_T$
over $\mu$, i.e.,
\begin{eqnarray}
P(T,\mu)=\int_{-\infty}^{\mu}n(T,\mu^\prime)d\mu^\prime.
\label{Pofmu}
\end{eqnarray}
Then we calculate the free-energy density $F/V$, the entropy density
$S/V$, and the energy density $E/V$ from the thermodynamic
relations
\begin{equation}
F=-PV+\mu nV\,,\quad
S=-\left.\frac{\partial F}{\partial
T}\right|_n\,,\quad\mbox{and}\quad
E=F+TS\,.
\label{thermodynamics}
\end{equation}

\section {Numerical results}
\label{sec:numerics}

\subsection {Density and the superfluid critical temperature}
\begin{figure}
\includegraphics[width=7.5cm]{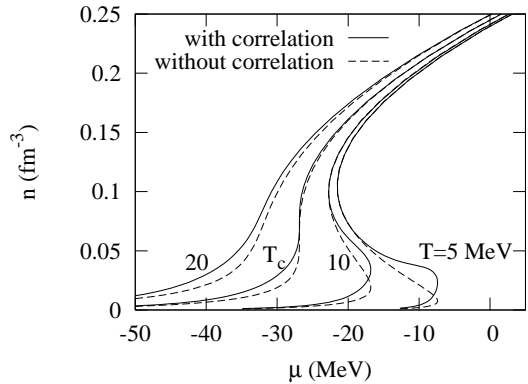}
\caption{The densities at $T = 20$, $15.9$, $10$, and $5$ MeV (from
  left to right) as functions of the chemical potential within Gogny
  HF (dashes) and with correlations (solid line). $T_c^\liqgas = 15.9$ MeV
  is the critical temperature for the liquid-gas phase transition.
\label{fig:nvsmu}}
\end{figure}

\begin{figure}
\includegraphics[width=7.5cm]{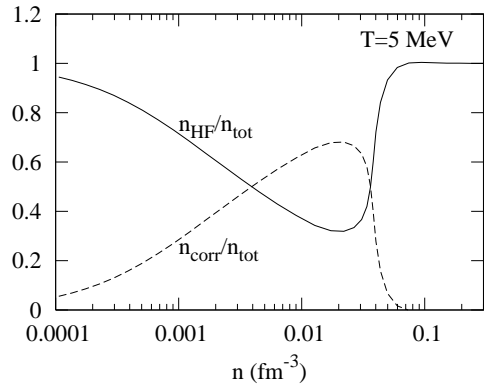}
\caption{HF and correlation contributions to the total density
  $n_\tot = n_\HF+n_\corr$ for $T = 5$ MeV.\label{fig:composition5}}
\end{figure}

\begin{figure}
\includegraphics[width=7.5cm]{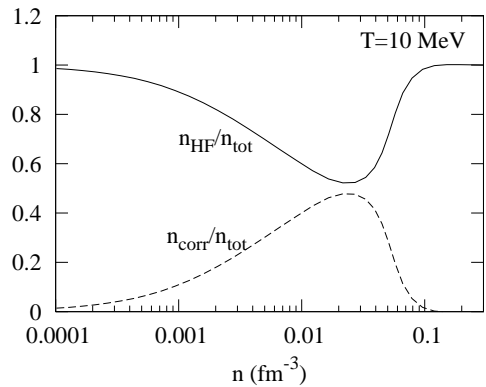}
\caption{Same as \Fig{fig:composition5}, but for $T = 10$
MeV.\label{fig:composition10}}
\end{figure}

\begin{figure}
\includegraphics[width=7.5cm]{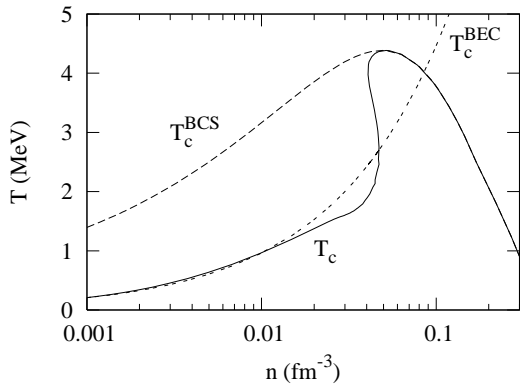}
\caption{Superfluid critical temperature as a function of the (total)
  density. The solid line is the full calculation, while the long
  dashes correspond to the BCS result. The short dashes show the
  critical temperature of Bose-Einstein condensation of a deuteron
  gas.\label{fig:superfluid}}
\end{figure}

We calculate the total density by numerically integrating \Eqs{nbound}
and (\ref{nscatt}). The results for the densities at different
temperatures as functions of the chemical potential%
\footnote{Strictly speaking, $n$ is not a function of $\mu$ since it
  is not single-valued, as will be discussed later. In practice, we
  generate the curves in \Figs{fig:nvsmu}-\ref{fig:entropy} by making
  a loop over the HF density and not over $\mu$.}
are shown in \Fig{fig:nvsmu}.
Comparing the results with correlations (solid lines) with the Gogny
HF results (dashed lines), one can see that, for a given chemical
potential, the correlations increase the densities. In the
high-density region, we notice that the results with and without
correlations converge to the same value, i.e., the correlations fade
away at high density, as this can be expected. For example, at $T=5$
MeV, the two results coincide starting from $n=0.07$ fm$^{-3}$. This
is a consequence of the Mott mechanism, which has been discussed at
length in \Ref{schmidt}. As mentioned above, the critical number
density where the bound state (at $K=0$) disappears is called Mott
density. When we change the temperature from 5 MeV to 10, 15.9, and 20
MeV, the Mott density changes from 0.07 fm$^{-3}$ to 0.12, 0.18, and
0.22 fm$^{-3}$. This means that the mean field approximation is valid
in the high density region. Below this region, the contribution of the
nucleon-nucleon correlations is important.

From this figure we also can see that when the temperature is less
than some critical value ($T_c^\liqgas=15.9$ MeV), the number density has
three values corresponding to one definite value of chemical
potential. This is a typical feature of the liquid-gas phase
transition in nuclear matter. We will discuss this phenomenon in
detail in the next subsection.

To see how large the correlation contribution to the density is, we
show the composition of the system at different temperatures in
\Figs{fig:composition5} and \ref{fig:composition10}.
Since the density ratios are shown as functions of the density and not
of the chemical potential, there are unique solutions even for
temperatures below $T_c^\liqgas$. In \Fig{fig:composition5} one can
see that at $T=5$ MeV the correlation contribution to the total
density is important at low density ($n<n_0/4$). At $n=0.02$
fm$^{-3}$, the correlated part is even larger than the HF part. This
means that most of the nucleons are in correlated pairs in this
density region. With increasing temperature, e.g., at $T = 10$ MeV as
shown in \Fig{fig:composition10}, the ratio of the correlated density
to the total density decreases, but the density region with sizeable
nucleon correlations is enlarged. Here we do not separate the
correlation contribution into bound and scattering state
contributions, since individually they are not very meaningful, as
discussed in \Ref{schmidt}. For instance, if the temperature is much
higher than the deuteron binding energy, the first term of the
scattering-state contribution (\ref{nscatt}) cancels almost exactly
the bound-state contribution (\ref{nbound}).

In the above calculation, when the temperature is below some critical
value, we get a divergence in the $T$ matrix. This pole corresponds to
the formation of Cooper pairs at high density and to Bose-Einstein
condensation of deuterons at low density. Below this critical
temperature $T_c$, the equations for the density of the system are
not applicable any more. In the superfluid phase, one would have to
include the nucleon pairing gap explicitly in the single-particle
Green's function (which then becomes a $2\times 2$ matrix in
Nambu-Gorkov space \cite{fetter}), which is beyond the scope of this
paper. However, we can determine the critical temperature of the
superfluid transition as the temperature where the $T$ matrix develops
a pole at zero total momentum ($K=0$) and at zero energy ($\omega =
0$). This is the well-known Thouless criterion \cite{thouless} for the
onset of superfluidity, coinciding with the BCS gap equation when the
gap $\Delta$ goes to zero:
\begin{eqnarray}
1-J_\alpha(K=0,\omega=0;T=T_c)=0.
\label{Thouless}
\end{eqnarray}
From this equation we get the critical temperature as a function of
the effective chemical potential. Using the relation between the
effective chemical potential and the number density, we obtain the
superfluid region beyond the BCS (mean field) result as shown in
\Fig{fig:superfluid}.
Qualitatively, this result is similar to the one in \cite{stein_nsr}
except that we have a lower critical temperature for the superfluid
phase transition. The maximum $T_c$ in \cite{stein_nsr} is 7.2 MeV at
$n=0.12$ fm$^{-3}$, while we have $T_c = 4.5$ MeV at $n=0.05$
fm$^{-3}$. The difference stems from the Gogny mean field, in
particular from the effective mass, which was neglected in
\Ref{stein_nsr}. One realizes that a $T_c$ of 4.5 MeV is still very
high, leading to a maximal gap of about 7 MeV, about three times as
much as the maximum value of the neutron-neutron gap in the spin
singlet channel. The reason clearly stems from the slightly stronger
attraction in the proton-neutron isoscalar channel. However, in finite
nuclei barely any enhancement of pairing in the $S=1$, $T=0$ channel
can be detected. Probably important screening is at work in that
channel. In nuclear matter, this has been investigated in
\Ref{cao}. The addition of screening effects is, however, beyond the
scope of this paper.

As mentioned above, $T_c$ as a function of $\mu$ coincides with the
BCS result. As a function of the density, the difference between the
results $T_c(n)$ with and without correlations comes only from the
different relations for $n$ as a function of $\mu$. Since the
correlation contribution to the density vanishes in the high density
region, the phase boundary coincides with the BCS curve (long dashed
line, which is obtained with $n_{HF}$ only). At very low density and
temperature, the main contribution to the density comes from the
deuteron bound state, as can be seen from \Eqs{ncontrib} and
(\ref{nbound}). Close to the Bose critical temperature, the Bose
distribution function in \Eq{nbound}) starts to diverge and,
therefore, dominates the whole expression for the density. Therefore
the superfluid critical temperature at low density coincides with the
critical temperature for Bose-Einstein condensation of a deuteron gas,
which is given by
\begin{equation}
T_c^\mathit{BEC} = \frac{\pi}{m}\left(\frac{n}
  {6\zeta(3/2)}\right)^{2/3}\,,
\end{equation}
(with $\zeta(3/2) = 2.612\dots$) and is shown as the short-dashed line
in \Fig{fig:superfluid}.

A surprising behavior of our result is that in the density region
between 0.04 fm$^{-3}$ and 0.05 fm$^{-3}$, \Eq{Thouless} for the
critical temperature has three solutions for one given density. This
behavior is not easy to understand from physical intuition. It seems
to be related to the effective mass, since it is absent in
\Ref{stein_nsr}. Anyway, as we will show in the next subsection, this
density region lies inside the unstable region of the liquid-gas phase
transition.

\subsection {Pressure and liquid-gas transition}
As it was shown in \Fig{fig:nvsmu}, there is a region of densities
where the chemical potential decreases with increasing density. This
is a typical feature of a liquid-gas phase transition. In order to
determine the boundary of this first-order phase transition, we need
the pressure. In principle, one can get the pressure as a function of
temperature and chemical potential, $P(T,\mu)$, from the number
density $n(T,\mu)$ by integration over the chemical potential $\mu$,
cf. \Eq{Pofmu}. However, since there is a first-order phase
transition, $n$ is not a single-valued function of $\mu$ any more. We
therefore transform the integral over $\mu$ into an integral over
$n_{HF}$:
\begin{equation}
P(T,n_\HF) = \int_0^{n_\HF(T,\mu)} n(T,n'_\HF)
  \left.\frac{\partial\mu}{\partial n'_\HF}\right|_T dn'_\HF\,.
\label{Pofn}
\end{equation}
Since $\mu$ is a single-valued function of $n_\HF$ (see dashed line in
\Fig{fig:nvsmu}), this integral is well defined. In this way we obtain
the pressure as a function of $n_\HF$, but neither $n_\HF$ nor $P$ are
single-valued functions of $\mu$.

\begin{figure}
\includegraphics[width=7.5cm]{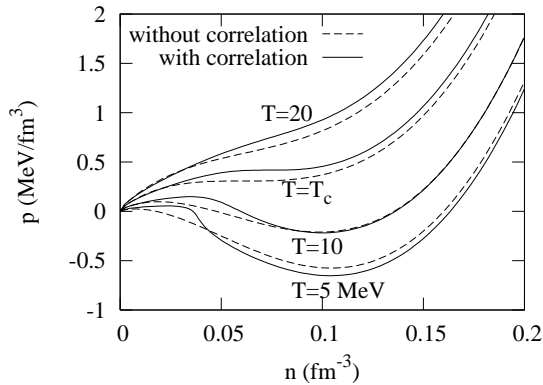} \caption{The pressure
  as a function of density at different
  temperatures. Solid lines: with correlations; dashed lines:
  mean-field results.
\label{fig:pressure}}
\end{figure}

If we plot the pressure as a function of the total density $n$ instead
of $n_\HF$, we get the results shown in \Fig{fig:pressure}.
Unfortunately, we cannot calculate the pressure for $T < 4.5$ MeV, at
least not at densities above $0.05$ fm$^{-3}$, because our method to
calculate the pressure at a given density $n$ necessitates the
calculation of all densities $n'< n$, i.e., including the density at
$n = 0.05$ fm$^{-3}$ where $T_c$ is maximum. For comparison, we also
give the results for the pressure within the mean-field approximation
(dashed lines in \Fig{fig:pressure}). As it can be seen, the main
effect of the nucleon-nucleon correlations is to increase the pressure
at very low densities. However, in the case $T = 5$ MeV shown in
\Fig{fig:pressure}, the pressure at higher densities is lower than the
HF result.

\begin{figure}
\includegraphics[width=7.5cm]{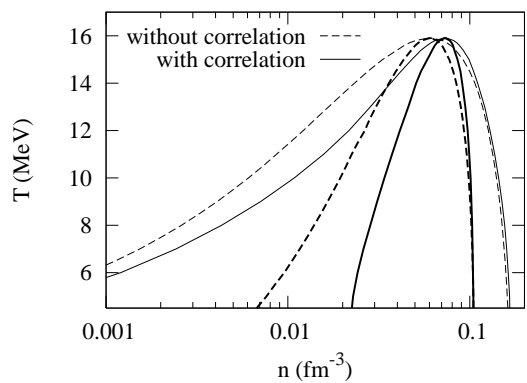}
\caption{The liquid-gas phase diagram as function of density and
temperature (for $T \geq 4.5$ MeV). The thin lines are the boundary of
the coexistence region, while the thick lines are the boundary of the
spinodal region. Solid lines: with correlations; dashed lines: mean
field results.
\label{fig:liquidgas}}
\end{figure}

Using the pressure, one can determine the coexistence region of the
liquid and gas phases of nuclear matter from the following conditions:
\begin{equation}
P(T,n_1) = P(T,n_2)\quad\mbox{and}\quad \mu(T,n_1)=\mu(T,n_2).
\end{equation}
The result is shown in \Fig{fig:liquidgas} as the thin solid line. At
the same time, we can determine the spinodal curve from the zeros of
$\partial P/\partial n$ (or, equivalently, of $\partial\mu/\partial
n$), which is shown as the thick solid line in \Fig{fig:liquidgas}. In
the region under the spinodal curve, the system cannot exist in a
homogeneous phase. In the region between the thin solid line and the
spinodal curve, the gas phase (left-hand part) or the liquid phase
(right-hand part) can exist as a metastable state. For comparison, the
corresponding mean-field results are presented in \Fig{fig:liquidgas}
as the dashed lines, which coincide with Fig. 6 of \Ref{ventura}.

Comparing the results with and without correlations, one can see that
the correlations decrease the phase-transition temperature in the
low-density region and reduce the unstable region of the liquid-gas
phase transition considerably. As mentioned in the introduction, this
is an expected result. In the high density region, the effect of the
correlations is almost negligible.

We can determine the critical temperature of the liquid-gas
transition, i.e., the maximum temperature of the coexistence and the
spinodal curves, from
\begin{equation}
\left.\frac{\partial P}{\partial n}\right|_{T_c^\liqgas}
  =\left.\frac{\partial^2 P}{\partial n^2}\right|_{T_c^\liqgas}=0,
\end{equation}
see \Fig{fig:pressure}. In this way, we obtain $T_c^\liqgas=15.9$ MeV,
which coincides with the mean-field result \cite{ventura,song}. The
fact that $T_c^\liqgas$ remains unchanged is an artifact of our
present approach to treat the correlation effects only at a
perturbative level, as explained in \Sec{sec:formalism}. As shown in
\Ref{roepke}, the inclusion of deuteron (and heavier) clusters should
reduce the liquid-gas critical temperature. We would have to do the
calculation more self-consistently in order to get a lower critical
temperature than the mean-field result.

\begin{figure}
\includegraphics[width=7.5cm]{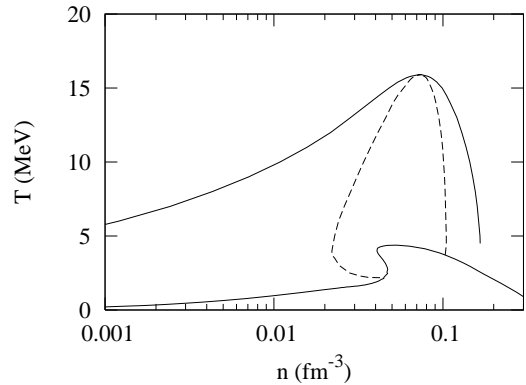}
\caption{Phase diagram combining the boundary of the superfluid phase
(lower curve), the liquid-gas coexistence region (upper curve), and
the spinodal line (dashed curve). The reason why the spinodal and
coexistence curves end at $T_c$ and $4.5$ MeV, respectively, is not
physical but it simply means that our model does not allow us to
compute them at lower temperatures (see text).
\label{fig:superfluidliquidgas}}
\end{figure}

In \Fig{fig:superfluidliquidgas},
the results of \Fig{fig:superfluid} for the superfluid critical
temperature $T_c$ (lower solid line) and \Fig{fig:liquidgas} for the
liquid-gas coexistence region (upper solid line) and the spinodal
instability region (dashed line) have been combined in a single phase
diagram. As explained above, we unfortunately cannot calculate the
liquid-gas coexistence curve for $T < 4.5$ MeV, but extrapolating the
solid curve to lower temperatures and remembering that at $T=0$ the
liquid phase gets stable at saturation density, it is clear that the
coexistence curve will cross the superfluid $T_c$ curve at $n\sim
n_0$, i.e., as one would expect, homogeneous nuclear matter with
pairing is stable above this density. From the results of
\Ref{stein_liquidgas} one can presume that the liquid-gas coexistence
region will be slightly reduced below the superfluid critical
temperature, but this effect should be almost negligible in the case
of symmetric nuclear matter considered here \cite{stein_liquidgas,su}.
At low densities, superfluid matter is never stable, because the
superfluid $T_c$ curve stays always below the coexistence curve.

The spinodal curve (dashed line) can be calculated until it reaches
the superfluid region. From this we see that superfluid nuclear matter
is metastable below $n\sim 0.045$ fm$^{-3}$ and above $n\sim 0.1$
fm$^{-3}$. Note that on the low-density side, the density region where
the gas phase is metastable is strongly increased by the correlations,
especially when we approach the superfluid transition
temperature. This confirms our expectation mentioned in the
introduction that the correlations have a stabilizing effect. However,
the BEC-BCS crossover lies still in the unstable region of the
liquid-gas phase transition.

\subsection{Energy and entropy}
\begin{figure}
\includegraphics[width=7.5cm]{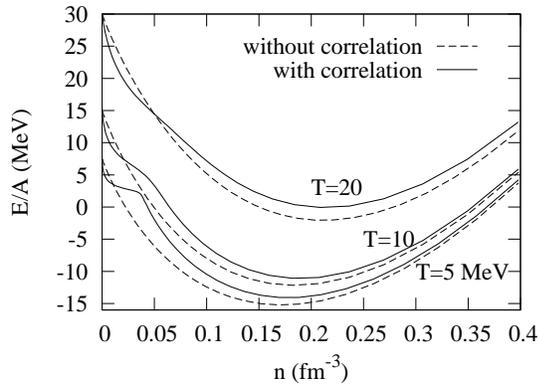}
\caption{Energy per nucleon as a function of density for different
temperatures. Solid lines: with correlations; dashed lines: mean field
results.\label{fig:energy}}
\end{figure}

\begin{figure}
\includegraphics[width=7.5cm]{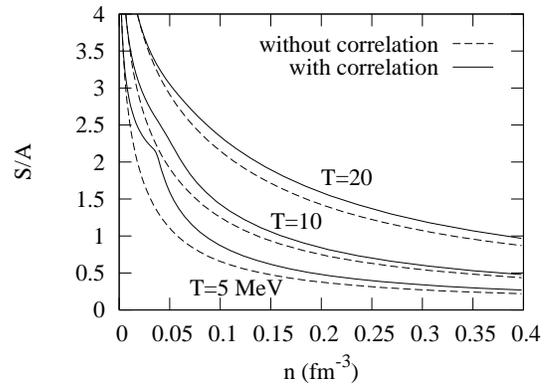}
\caption{Entropy per nucleon as a function of density for different
temperatures. Solid lines: with correlations; dashed lines: mean field
results.\label{fig:entropy}}
\end{figure}

The energy and the entropy can be obtained from the pressure with the
help of the thermodynamic relations (\ref{thermodynamics}). Results
for the energy per nucleon, $E/A$, and for the entropy per nucleon,
$S/A$, for different temperatures are shown in \Figs{fig:energy} and
\ref{fig:entropy}.
The corresponding mean-field results (dashed lines) are also shown for
comparison. The results shown in \Fig{fig:energy} indicate that, for
fixed temperature, the correlations shift the minimum of the energy
per nucleon to slightly higher densities. Fortunately the change is
very small, because otherwise we would have to readjust the parameters
of the Gogny force, which gives the right saturation density and
energy at zero temperature without correlations.

In the low-density region, where the deuterons and the nucleon-nucleon
scattering states dominate, the energy per nucleon is lower than that
the HF result. When the density is high, the correlation effect goes
to zero and the energy per nucleon gets close to the mean-field
result.

When the density approaches zero, both results go to the classical
value of an ideal gas of nucleons,
\begin{equation}
\lim_{n\to 0}E/A = \tfrac{3}{2} T.
\label{elimit}
\end{equation}
This is not surprising, since even the lowest temperature considered
here, $T = 5$ MeV, is still much higher than the deuteron binding
energy so that almost all deuterons will be dissociated. However, the
result (\ref{elimit}) is also found at temperatures much lower than
the deuteron binding energy. This is because, at finite temperature,
the deuterons are always dissolved in the low-density limit. This is a
consequence of the mass-action law and can easily be understood as
follows: At low density, the chemical potential of the nucleons,
$\mu$, gets strongly negative, $\mu \ll -T$. The chemical potential of
the deuterons is $2\mu$, which is even more negative. So the nucleon
density $\propto e^{\mu/T}$ is much larger than the deuteron density
$\propto e^{2\mu/T}$. Only at zero temperature, where the system
remains a deuteron BEC at arbitrarily low densities, the energy per
nucleon approaches $-1.12$ MeV (half the deuteron binding energy) in
the limit $n\to 0$ \cite{baldo}.

The results for the entropy (cf. \Fig{fig:entropy}) have been
calculated from \Eq{thermodynamics} and show that, for fixed
temperature, the entropy per nucleon decreases with increasing
density. In the zero-density limit, the entropy per nucleon increases
logarithmically, in agreement with the result for a classical ideal
nucleon gas. As is clear from the discussion above, the correlations
do not change this asymptotic behavior. At slightly larger values of
the density, the correlations tend to increase the entropy.

\section {Summary}
In this paper, we discussed the effect of pair correlations beyond the
mean-field approximation in symmetric nuclear matter above the
superfluid critical temperature. We include the effects of
non-condensed pairs (deuterons) as well as the contribution of
scattering states. For the mean field, we use the Gogny effective
interaction in order to get the right saturation properties of nuclear
matter.

Starting from the single-particle Green's function within the Gogny HF
approximation, we include the correlations in a perturbative way by
considering in addition to the HF Green's function the diagram with
one self-energy insertion, the self-energy being calculated in ladder
approximation. This approximation scheme is analogous to the
well-known NSR approach. However, in
order to avoid double counting of the quasiparticle energy shift which
is already accounted for by the Gogny mean field, we have to subtract
the self-energy at the quasiparticle energy. This leads finally to the
same formula for the density in terms of the in-medium scattering
phase shifts as given in \Ref{stein_nsr}. We use a separable
Yamaguchi potential in order to get an analytical formula for the
in-medium $T$ matrix and the phase shifts.

Evaluating numerically these formulas for the density, we discussed
the different density contributions in hot and dense nuclear matter
and found that the nucleon-nucleon correlations are important in the
low-temperature and low-density region ($n<n_0$). The correlation
effect on the superfluid critical temperature was discussed. The
result interpolates between the critical temperature for Bose-Einstein
condensation at low density and the BCS critical temperature at high
density. We found that the maximum of the superfluid critical
temperature decreases from $7.2$ MeV (the value given in
\Ref{stein_nsr}) to $4.5$ MeV when the effective mass $m^*$ due to the
Gogny mean field is taken into account.

Then we studied the liquid-gas phase transition in hot and dense
nuclear matter with the help of the pressure calculated from the
density. Especially at low density, we found that the boundaries of
the coexistence and spinodal regions of the phase transition are
shifted by the pair correlations. As we expected, the stable and
metastable regions of the gas phase are strongly enlarged. In
particular near the superfluid transition temperature, the gas phase
stays metastable up to much higher densities if the correlations are
taken into account. However, the correlations are not strong enough to
suppress the liquid-gas transition. This could have been anticipated
from the fact that the liquid-gas critical temperature is much higher
than the superfluid one \cite{stein_liquidgas,su}. Because of our
perturbative treatment of the correlations, the critical temperature
of the liquid-gas transition remains the same as within the mean field
approximation.

Finally, we calculated the energy and entropy of nuclear matter from
thermodynamic relations. The nucleon-nucleon correlations decrease the
energy per nucleon in the low density region but increase it at high
density. For the entropy, the correlations always give a positive
contribution.

As mentioned before, our result for the critical temperature of
liquid-gas phase transition is not affected by the pair
correlations because they are treated only perturbatively. One
should improve this by taking the correlations into account
self-consistently. Then the correlations will have some effect on
the HF field and the critical temperature will change. The
saturation point of nuclear matter, given correctly by the Gogny
interaction within the HF approximation, may be changed,
necessitating a readjustment of the parameters of the Gogny force.

Our equation of state is only valid for temperatures and densities
above the superfluid critical temperature. In order to get a result
which is valid in the whole temperature and density plane, one should
introduce the pairing gap $\Delta$ into the single-particle Green's
functions. Some work in this direction has been done for nuclear
matter \cite{bozek}, and quite elaborate theories have been developed
for the BEC-BCS crossover in ultracold atomic Fermi gases
\cite{perali}. We leave this for future study. Another important
extension of the present work would be to consider the case of
asymmetric nuclear matter and neutron matter, since these are of great
importance for the study of neutron stars and their formation.

\acknowledgments{We thank G. R\"opke and J. Margueron for discussions
and helpful remarks. M.J. acknowledges financial support from
Universit\'e Paris-Sud 11 and from CNRS during his postdoctoral stays
at Institut de Physique Nucl\'eaire d'Orsay, where most of this work
has been done. This work was supported in part by NSFC (grants
10805023 and 10975060) and by ANR (project NEXEN).}
\appendix
\section*{Appendix: Derivation of the density formulas}
In this Appendix we give a more transparent derivation of the density
formulas (\ref{ncontrib}), (\ref{nbound}), and (\ref{nscatt}), which
were initially derived in \Refs{zimmermann,schmidt}. For better
readability, we will not write out the sum over $\alpha =$ $^3S_1$,
$^1S_0$ and suppress the index $\alpha$ in this appendix.

Let us recall the spectral representation of the $T$ matrix,
\begin{equation}
T(k,k',K,\omega) = V(k,k')-\int\frac{d\omega'}{\pi} \frac{\Im
  T(k,k',K,\omega')}{\omega-\omega'+i0}\,,
\label{Tspectral}
\end{equation}
where $\omega$ can be real or complex. Analogous dispersion relations
exist for the self-energy $\Sigma(p,\omega)$ and for the two-particle
propagator $J(K,\omega)$ defined in \Eq{Jalpha}. Using \Eq{Tspectral},
one can evaluate the frequency sum in \Eq{SigmaMatsubara}, and one
obtains the well-known expression for the imaginary part of the
self-energy:
\begin{multline}
\Im\Sigma(p,\omega) = \frac{3}{2} \int\frac{d^3p'}{(2\pi)^3}\Im
  T(k,k,K, \omega+\xi_{p'})\\
  \times[f(\xi_{p'})+g(\omega+\xi_{p'})]\,.
\label{ImSigma}
\end{multline}
where $\vec{k}$ and $\vec{K}$ are the relative and total momenta as
defined below \Eq{SigmaMatsubara}.

The correlation correction to the density is given by
\begin{equation}
n_\corr = -4 T\sum_{n}\int\frac{d^3p}{(2\pi)^3}
  \frac{\Sigma(p,i\omega_n)-\Re\Sigma(p,\xi_p)}{(i\omega_n-\xi_p)^2}\,.
\end{equation}
If we use the spectral representation of $\Sigma$, the frequency sum
can be evaluated with the result
\begin{equation}
n_\corr = -4\int \frac{d^3p}{(2\pi)^3}
 \mathcal{P}\!\!\int\frac{d\omega}{\pi} \Im \Sigma(p,\omega)
 \frac{f(\omega)-f(\xi_p)}{(\omega-\xi_p)^2}\,,
\end{equation}
where $\mathcal{P}$ denotes the principal value. Inserting
\Eq{ImSigma} into this expression, one obtains with the help of the
relation $f(\xi_p)f(\xi_{p'}) = g(\xi_p+\xi_{p'})
[1-f(\xi_p)-f(\xi_{p'})]$ and after some transformations
\begin{multline}
n_\corr = -6 \int \frac{d^3p\,d^3p'}{(2\pi)^6}
  \mathcal{P}\!\!\int\frac{d\omega}{\pi} \Im T(k,k,K,\omega)\\
  \times (1-f(\xi_p)-f(\xi_{p'}))
  \frac{g(\omega)-g(\xi_p+\xi_{p'})}{(\omega-\xi_p-\xi_{p'})^2}\,.
\end{multline}
The next step is to introduce the new variable
$\omega'=\xi_p+\xi_{p'}$ and to replace the integral over $p'$
by an integral over $\omega'$. Then, using the imaginary parts of
\Eqs{Tmatrix} and (\ref{Jalpha}), one can show that the resulting
expression for $n_\corr$ can be rewritten as
\begin{multline}
n_\corr = 6 \int \frac{d^3K}{(2\pi)^3}
  \mathcal{P}\!\!\int\frac{d\omega\,d\omega'}{\pi^2}
  \Im\frac{1}{1-J(K,\omega)}\\
  \times \Im J(K,\omega')
  \frac{g(\omega)-g(\omega')}{(\omega-\omega')^2}\,.
\end{multline}
With the help of the dispersion relations for the real parts, this
expression can be further reduced to
\begin{multline}
n_\corr = 6 \int \frac{d^3K}{(2\pi)^3} \int
  \frac{d\omega}{\pi} g(\omega)
  \left(\Im \frac{1}{1-J}\frac{d}{d\omega}\Re J\right.\\
    \left.-\Im J\frac{d}{d\omega} \Re\frac{1}{1-J}\right)
\label{ncorr1}
\end{multline}
(the arguments of $J(K,\omega)$ have been suppressed for brevity).
In order to express everything in terms of the in-medium
scattering phase shifts $\delta = -\Im \ln(1-J)$, we notice that
\begin{gather}
\frac{d\delta}{d\omega} = \Im \frac{1}{1-J}\frac{d}{d\omega}\Re J
  +\Re \frac{1}{1-J}\frac{d}{d\omega}\Im J\,,\\
\Im J\Re\frac{1}{1-J} = \sin\delta\cos\delta = \tfrac{1}{2}\sin2\delta\,.
\end{gather}
With these relations, \Eq{ncorr1} can be rewritten as
\begin{equation}
n_\corr = 6 \int \frac{d^3K}{(2\pi)^3} \int \frac{d\omega}{\pi}
  g(\omega) \frac{d}{d\omega} \left(\delta -
  \tfrac{1}{2}\sin2\delta\right)\,.
\end{equation}
The final step is to integrate by parts over $\omega$ and to separate
in the resulting integral the contributions of $\omega > \omega_0(K)$
(scattering-state contribution $n_\scatt$) and $\omega < \omega_0(K)$
(bound-state contribution $n_\bound$). The latter reduces to
\Eq{nbound} since the phase shift below threshold is (see also
Fig. 7 of \Ref{nsr})
\begin{equation}
\delta(K,\omega < \omega_0(K)) = \begin{cases}
  0, &\mbox{if $K < K_\Mott$,}\\
  \pi\theta(\omega-\omega_b(K)), &\mbox{if $K > K_\Mott$.}
\end{cases}
\end{equation}

\end{document}